\documentclass[conference]{IEEEtran}
\usepackage{cite}
\usepackage[table, dvipsnames]{xcolor}
\usepackage{booktabs, multicol, multirow}
\usepackage{graphicx, amsfonts,amsmath , boldline, epstopdf,amssymb }
\usepackage{tabu, multirow,multicol ,booktabs ,setspace ,algcompatible ,mathrsfs,dsfont, array, boldline, makecell, booktabs }
\usepackage{url}
\usepackage{tikz}
\usepackage{epstopdf}
\usepackage{subcaption} 
\usepackage{float}      
\usepackage{bm}
\usepackage{xcolor}     
\usepackage[font=small, labelfont=bf]{caption}
\usepackage{adjustbox, stackengine, color, colortbl, epsfig,cite, cases, enumitem, mathtools, tikz, verbatim}
\newcolumntype{?}{!{\vrule width 1.2pt}} 
\usepackage[caption = false]{subfig}

\usepackage[utf8]{inputenc}
\usepackage{makecell}
\usepackage{tikz}

\usepackage[utf8]{inputenc}
\usetikzlibrary{arrows.meta, positioning}
\usepackage[utf8]{inputenc}
\usepackage{placeins}
\usetikzlibrary{tikzmark}

\newcommand*{\Rectangle}[1][]{%
  \tikz[baseline=-0.6ex]\node[draw, minimum width=0.5cm,minimum height=0.05cm, fill={#1}]{};
}
\tikzset{mycircled/.style={circle,draw,inner sep=0.1em,line width=0.1em}}

\definecolor{bb}{rgb}{0.2941, 0.5447, 0.7494}
\definecolor{orange}{rgb}{1, 0.347, 0}
\definecolor{greenJ}{rgb}{0, 0.6590, 0.42}
\definecolor{zereshk}{rgb}{0.588,0,0.098}
\definecolor{Maroon}{rgb}{0.502, 0, 0}
\definecolor{Brown}{rgb}{0.588, 0.294, 0}
\definecolor{Olive}{rgb}{0.502, 0.502, 0}
\definecolor{Navy}{rgb}{0, 0, 0.502}
\definecolor{Orange}{rgb}{1,0.647,0}
\definecolor{Yellow}{rgb}{0.502, 1, 0}
\definecolor{Green}{rgb}{0, 0.502,  0}
\definecolor{Blue}{rgb}{0, 0,0.761}
\definecolor{Lime}{rgb}{0.196, 0.804, 0.196}
\definecolor{Purple}{rgb}{0.502,0,0.502}
\definecolor{Violet}{rgb}{0.561,0,1}
\definecolor{Magneta}{rgb}{1,0,1}
\definecolor{Red}{rgb}{1,0,0}
\definecolor{Abi}{rgb}{0.059, 0.322, 0.729}
\definecolor{Gray}{rgb}{0.502, 0.502, 0.502}
\definecolor{um}{rgb}{0.0824, 0.1294, 0.4196}
\definecolor{abikam}{rgb}{0.51, 0.93,  0.992}
\definecolor{abizeyad}{rgb}{0.16, 0.573,0.761}
\definecolor{rr}{rgb}{0.9047, 0.1918, 0.1988}

\definecolor{orange2}{rgb}{0.85,0.33,0.10}
\definecolor{yellow4}{rgb}{0.93,0.69,0.13}
\definecolor{purple6}{rgb}{0.49,0.18,0.56}
\definecolor{green32}{rgb}{0.47,0.67,0.19}
\definecolor{blue13}{rgb}{0.30,0.75,0.93}
\definecolor{zereshk16}{rgb}{0.64,0.08,0.18}
\definecolor{orange25}{rgb}{0.93,0.69,0.13}
\definecolor{gray27}{rgb}{0.0824, 0.1294, 0.4196}

\definecolor{storageBlue}{rgb}{0, 0.4470, 0.7410}   
\definecolor{storageOrange}{rgb}{0.8500, 0.3250, 0.0980} 
\definecolor{storageYellow}{rgb}{0.9290, 0.6940, 0.1250} 
\definecolor{storagePurple}{rgb}{0.4940, 0.1840, 0.5560} 
\definecolor{storageGreen}{rgb}{0.4660, 0.6740, 0.1880} 

\begin{document}

\title{Robust Optimal Power Flow  Against  Adversarial Attacks: A Tri-Level Optimization Approach}
 
\allowdisplaybreaks
\IEEEaftertitletext{\vspace{-2.2\baselineskip}}

\normalsize{
\author{
\IEEEauthorblockN{Saman Mazaheri Khamaneh,  \textit{IEEE   Student Member}, Tong Wu, \textit{IEEE Member}}
\IEEEauthorblockA{Department of Electrical and Computer Engineering, University of Central Florida, Orlando, FL, USA\\
smazaheri@ucf.edu, tong.wu@ucf.edu}
 \vspace{-2.8\baselineskip}
}

\maketitle
 
\thispagestyle{empty}
\pagestyle{empty}

\begin{abstract}
In power systems, unpredictable events like extreme weather, equipment failures, and cyberattacks present significant challenges to ensuring safety and reliability. Ensuring resilience in the face of these uncertainties is crucial for reliable and efficient operations. This paper presents a tri-level optimization approach for robust power system operations that effectively address worst-case attacks.  The first stage focuses on optimizing economic dispatch under normal operating conditions, aiming to minimize generation costs while maintaining the supply-demand balance. The second stage introduces an adversarial attack model, identifying worst-case scenarios that maximize the system's vulnerability by targeting distributed generation (DG). In the third stage, mitigation strategies are developed using fast-response energy storage systems (ESS) to minimize disruptions caused by these attacks. By integrating economic dispatch, vulnerability assessment, and mitigation into a unified framework, this approach provides a robust solution for enhancing power system resilience and safety against evolving adversarial threats.
The approach is validated using the IEEE-33 node distribution system to demonstrate its effectiveness in achieving both cost efficiency and system resilience.

\end{abstract}

\begin{IEEEkeywords}
Tri-level optimization, adversarial attacks, attack mitigation, power system resilience
\end{IEEEkeywords}

\allowdisplaybreaks
\section{Introduction}
\subsection{Background}


Power systems are essential for providing energy to modern society’s daily activities, industries, and services. However, they face increasing adversarial threats that challenge reliable operations. The integration of distributed energy resources (DERs) like wind and solar, though beneficial for sustainability, adds vulnerability due to inherent uncertainty, which complicates system stability and opens potential weaknesses for exploitation  \cite{wu2021deep}. The growing risk of cyberattacks on critical infrastructure further threatens the control, communication, and integrity of power systems \cite{yu2023survey, wu2022reinforcement}. Combined with extreme weather events like hurricanes, these threats pose severe risks of equipment failures and large-scale disruptions to energy delivery \cite{wang2015research}.

To address adversarial attacks on power systems, methods like security-constrained optimal power flow (OPF) provide foundational optimization approaches \cite{yu2023survey}. These techniques help develop defensive strategies to maintain system performance under challenging conditions. However, they often assume that attackers lack detailed knowledge of the system, potentially underestimating worst-case scenarios \cite{zhao2019resilient}. In reality, attackers with insights into system configurations and vulnerabilities may exploit weaknesses that go beyond the expected range of disruptions considered by traditional methods.

\subsection{Related Work and Our Contributions}
Robust minimax optimization has been widely explored in power system planning and operations to address worst-case scenarios, proving effective in mitigating extreme weather impacts, renewable variability, load fluctuations, and other uncertainties \cite{su2024review}. For example, \cite{yuan2016robust} developed a min-max framework for resilient distribution planning against natural disasters, and \cite{yan2019robust} applied it to battery storage systems for congestion management, though their model lacked real-time adaptability. Similarly, \cite{surani2024competitive} proposed a two-stage min-max approach to address adversarial attacks on the EV charging market, impacting power system costs. However, these methods tend to be overly conservative, often leading to excessive reserve allocation \cite{surani2024competitive}, and they overlook the differing response times of power system devices, particularly the rapid response of battery systems in mitigating cyberattack impacts.


To address the challenges outlined, we introduce a tri-level optimization framework for robust optimal power flow designed to counter adversarial attacks. The primary contributions of this paper are as follows:

\begin{itemize}
\item We formulate the robust OPF problem as a three-stage process, encompassing (i) hourly economic dispatch scheduling, (ii) assessment of potential worst-case adversarial threats based on the current system state, and (iii) mitigation using fast-response energy storage systems (ESSs). Additionally, the state of charge (SOC) of the ESS is included in the model to optimize renewable energy use for charging, ensuring sufficient power availability for future needs.
\item We design a novel tri-level optimization structure that sequentially integrates system operation, worst-case adversarial threat assessment, and mitigation strategies. This approach ensures that the mitigation stage can effectively address potential worst-case adversarial threats by targeting the most vulnerable devices and amplifying constraint violations based on real-time system conditions. In this way, system resilience is maximized by neutralizing potential attacks, as typical threats are likely to be weaker than the anticipated worst-case scenarios.
\end{itemize}

Numerical case studies show that worst-case attacks can severely disrupt power systems; however, our mitigation strategy can quickly restore system safety and enhance power system resilience.
\begin{figure}[htpb!]
    \centering
    \begin{tikzpicture}[
        node distance=2.0cm, auto, >=stealth,
        block/.style = {rectangle, draw, fill=cyan!30, text width=9em, text centered, rounded corners, minimum height=3em, font=\scriptsize},
        block_wide/.style = {rectangle, draw, fill=orange!30, text width=10em, text centered, rounded corners, minimum height=3em, font=\scriptsize},
        line/.style = {draw, thick, ->, shorten >=3pt, color=blue},
    ]
    \node [block] (opf) {
        \textbf{Stage 1: Distribution System OPF} \par 
        Minimize cost \par 
        \textbf{Equation:} $\min_{x \in \mathcal{X}} \ell_{\mathrm{base}}(x)$ \par 
        \textbf{s.t.} $f_{\text{base}}(x, v) = 0$, $g_{\text{base}}(x, v) \leq 0$
    };
    \node [block_wide, minimum height=2.5em, below left of=opf, xshift=-1.3cm, yshift=-1.4cm] (attack) {
        \textbf{Stage 2: Adversarial Attack} \par 
        Maximize loss \par 
        \textbf{Equation:} $\max_{y \in \mathcal{Y}} \ell_{\mathrm{cont}}(x^*, y) + \inf[g_{\mathrm{cont}}(x^*, y)]$ \par 
        \textbf{s.t.} $f_{\mathrm{cont}}(x^*, y)  = 0$
    };
    \node [rectangle, draw, fill=teal!30, text width=9em, text centered, rounded corners, minimum height=3em, font=\scriptsize, below right of=opf, xshift=1cm, yshift=-1.4cm] (mitigation) {
        \textbf{Stage 3: Attack Mitigation} \par 
        Mitigate violations \par 
        \textbf{Equation:} $\min_{z \in \mathcal{Z}} \ell(x^*, y^*, z) + \sup[g(x^*, y^*, z)]$ \par 
        \textbf{s.t.} $f(x^*, y^*, z) = 0$, $g(x^*, y^*, z) \leq 0$
    };

    \path [line,  shorten >=2.5pt, shorten <=2.5pt] (opf) -- node[midway, left, align=center, font=\scriptsize, color=blue, xshift=-2mm, yshift= 1mm] {Generation \\ Dispatch} (attack);
   \path [line,  shorten >=2.5pt, shorten <=2.5pt] (mitigation) -- node[midway, right, align=center, font=\scriptsize, color=blue, xshift=2mm, yshift= 1mm] {System \\Resilience} (opf);
    \path [line, shorten >=2.5pt, shorten <=2.5pt] (attack) -- node[midway, above, align=center, font=\scriptsize, color=blue] {Worst-case \\ Attack} (mitigation);
    \end{tikzpicture}
    \caption{Methodology overview of the proposed tri-level optimization framework.}
    \label{fig:overview}
    \vspace{-0.6cm}
\end{figure}
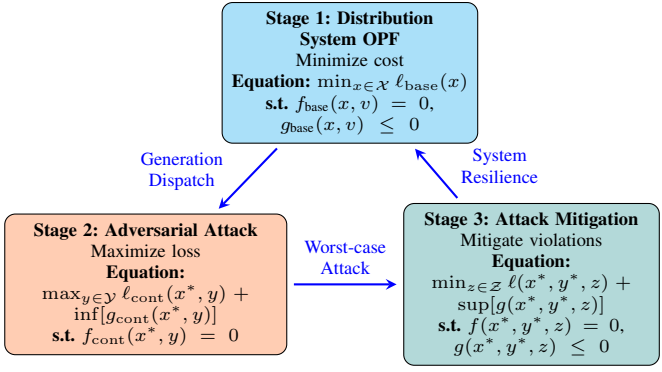

\section{Problem Formulation and Methodology} \label{sec:Gen_formula}
In this section, before presenting the detailed formulation of our proposed approach, we first introduce the generic framework of the tri-level optimization problem and discuss how to model the attacker-defender scenario within this framework as follows:
\begin{subequations} \label{mod:Gen}
    \begin{align}
&\min_{x \in \mathcal{X}} \max_{y \in \mathcal{Y}} \min_{z \in \mathcal{Z}} &&\ell(x, y, z)
\\ &\text{s.t.} && f(x, y, z) = 0, \label{con:equalit_constraint}
\\ &&& g(x, y, z) \leq 0, \label{con:non-equalit}
\end{align}
\end{subequations}
where the minimizer variables, such as $x \in \mathcal{X}$, represent the dispatched power generation by the grid's generators. In the second stage, the attacker variable $y \in \mathcal{Y}$ aims to maximize contingencies involving outages in both transmission lines and generators. In the third stage, we introduce the variable $z \in \mathcal{Z}$, representing the defender, which seeks to counter the attack using fast-response devices such as smart inverters and  ESSs.

Fig. \ref{fig:overview} illustrates the comprehensive methodology of our proposed model. Stage 1 focuses on economically dispatching generators under base conditions, optimizing costs while adhering to power flow and boundary constraints. Stage 2 assesses vulnerabilities by identifying the worst-case contingency based on the current power system state from Stage 1 to evaluate potential threats. Stage 3 determines optimal mitigation strategies, such as deploying ESSs, to provide fast-response solutions and enhance system resilience against identified threats. After implementing these mitigation actions, the system progresses to the next time point, incorporating new demand values for the subsequent iteration of Stage 1.

\subsection{Stage 1:OPF under Normal Operating Conditions} \label{sec: OPF_normal_condition}
In Fig. \ref{fig:overview}, Stage 1 solves the general optimization problem under normal conditions (base case) to minimize operational costs while meeting equality and inequality constraints. This establishes baseline performance for resilience analysis in later stages.
For example, the OPF problem illustrates this optimization. The OPF model includes equality constraints for power flow equations and inequality constraints for system limits, such as generation capacity, voltage stability, and line flows, ensuring safe operation. 
In model \eqref{mod:DSO}, the active power of DGs and substation generation (\(\boldsymbol{p}^\mathrm{g}\)), and the reactive power (\(\boldsymbol{q}^{\mathrm{g}}\)), are the dispatch variables.  Each node $i \in \mathcal{I}$ represents a network node, and each power flow line $l \in \mathcal{L}$ connects nodes, with FN$_l$ as the starting node and TN$_l$ as the ending node. Power system state values include active power flow  ${pf}_l$, reactive power flow  ${qf}_l$  on line $l$, and squared node voltage magnitude  ${v}$. The LF (or TF) matrix represents the incidence relationship between lines and nodes. Each element has a value of 1 if line \( l \) originates from (or terminates at) node \( i \), and zero otherwise.
\begin{subequations}
 \label{mod:DSO}
    \begin{align}
& \min_{\bm{x}}
        && \sum_{i \in \mathcal{I}^{\mathrm{g}} } \sum_{t \in \mathcal{T}} C_{i,t}^{g} p_{i,t}^{\mathrm{g}}    \label{obj:DSO} \\
        &\text{s.t.} && \sum_{l \in \mathcal{L}} pf_{l,t} \cdot {\mathrm{LT}_{l,i}} - \sum_{l \in \mathcal{L}} pf_{l,t} \cdot {\mathrm{LF}_{l,i}}     = P_{i,t}^d - p_{i,t}^{\mathrm{g}},   \label{cons:DSO_P_flow}\\
        & &&\sum_{l \in \mathcal{L}} qf_{l,t} \cdot {\mathrm{LT}_{l,i}} - \sum_{l \in \mathcal{L}} qf_{l,t} \cdot {\mathrm{LF}_{l, i}}   = Q_{i,t}^d - q_{i,t}^{\mathrm{g}},  \label{cons:DSO_Q_flow}\\
        & &&  PF_{l}^{\mathrm{min}} \leq pf_{l,t}  \leq  PF_{l}^{\mathrm{max}},   QF_{l}^{\mathrm{min}} \leq qf_{l,t}  \leq  QF_{l}^{\mathrm{max}}, \label{cons:DSO_line_stat}\\
        & && v_{\mathrm{FN}_{l},t}-v_{\mathrm{TN}_l,t} = 2 \cdot(r_{l} \cdot pf_{l,t} + x_{l} \cdot qf_{l,t}),  \label{cons:DSO_voltage} \\
        &&&   p^{g, \mathrm{min}}_i \leq p_{i,t}^{g} \leq  p^{g, \mathrm{max}}_i, q^{g, \mathrm{min}}_i \leq q_{i,t}^{g} \leq  q^{g, \mathrm{max}}_i, 
        \label{cons:DG_NA_gen_bound}\\
        & && (V^{\mathrm{min}}_i)^2 \leq v_{i,t} \leq (V^{\mathrm{max}}_i)^2, \label{cons:DSO_voltage_bounds}\\
        &&& \forall i\in\mathcal{I}, \forall l\in\mathcal{L}, \forall t\in\mathcal{T} \notag
    \end{align}
\end{subequations}
 where \( p^{g, \mathrm{min}}_i = p^{g, \mathrm{max}}_i = q^{g, \mathrm{min}}_i = q^{g, \mathrm{max}}_i = 0 \) for \( i \in \mathcal{I} \setminus \mathcal{I}^g \), ensuring no generation at non-generator nodes.
The objective function in \eqref{obj:DSO} minimizes the total generation cost, including both the cost of DG active power generation and power purchased from the substation. Constraint \eqref{cons:DSO_P_flow} ensures active power balance at each node, while constraint \eqref{cons:DSO_Q_flow} maintains reactive power balance.  Constraint \eqref{cons:DSO_line_stat} enforces line flow limits, ensuring that the active (and reactive) power flows on each line remain within the minimum \( PF_{l}^{\mathrm{min}} \) (and \( QF_{l}^{\mathrm{min}} \)) and maximum \( PF_{l}^{\mathrm{max}} \) (and \( QF_{l}^{\mathrm{max}} \)) limits. Voltage differences across lines are captured in \eqref{cons:DSO_voltage}, incorporating line resistance \( r_l \), reactance \( x_l \), and power flows. Constraint \eqref{cons:DSO_voltage_bounds} keeps node voltages within \( V^{\mathrm{min}}_i \) and \( V^{\mathrm{max}}_i \), and \eqref{cons:DG_NA_gen_bound} bounds active and reactive power generation. These constraints follow the Linearized Dist-Flow equations \cite{baran1989network, coffrin2016strengthening}.


\subsection{Stage 2: Adversarial Attack Assessment (AAA) Model}
After solving the OPF in Stage 1, the resulting power system states, along with the output of inertia-based generators, denoted by \( {x}^* \), remain unchanged as long as demand does not change. Given this current state, we consider the possibility of an adversarial attack intended to disrupt these states. This adversarial attack is represented by the primary contentious variable \( y_j \in [0,1] \), where \( j \) denotes a power generation outage. This variable is defined as follows:
\begin{equation}
    \label{mod:attack_var}
y_j = 
\begin{cases}
1 \text{ (or 0)}, & \text{if outage } j \text{ is fully active (or inactive)}, \\
\alpha_j \in (0, 1), & \text{if outage } j \text{ is partially active at level } \alpha_j.
\end{cases}
\end{equation}

In Stage 2 of Fig. \ref{fig:overview}, the adversarial attack maximizes the contingency loss function by violating system constraints to model a worst-case scenario, identifying vulnerabilities, and assessing impacts. The objective is to determine the worst-case contingencies with up to \( k \) simultaneous attacks on generation units. The first term, \( \ell_{\mathrm{cont}}(x^*, y) \), represents the cost of active power generation from attackable DGs, where the generation values \( x^* \) are optimally obtained from Stage 1 (model \eqref{mod:DSO}), and power purchased from the substation. The second term, \( \inf[g_{\mathrm{cont}}(x^*, y)] \), corresponds to \( \inf_{k=1,\dots,4, \forall l, \forall t} \Phi^k_{l,t} + \inf_{k=5,6, \forall i, \forall t} \Phi^k_{i,t} \) in \eqref{mod:Detailed_attack}. Here, the infimum is used to force even the safest constraints—those furthest from their limits—into violation, maximizing the number of constraints pushed into infeasibility and driving the system toward instability.

By maximizing this objective, the formulation identifies the combination of attacked DGs and constraint violations that yield the highest operational cost and lowest system stability, thereby revealing critical vulnerabilities within the power network, which can be expressed as:
\begin{subequations}
    \label{mod:Detailed_attack}
    \begin{align}
        & \max_{{y}, {p}^g_{{sub}}} 
        && \sum_{i \in \mathcal{I}^{g}_a} \sum_{t \in \mathcal{T}} C_{i,t}^{g} p_{i,t}^{g*} (1- y_{i,t}) +  \sum_{t \in \mathcal{T}} C^g_{\mathrm{sub},t}  p^g_{\mathrm{sub},t}  \nonumber \\
        &&& + \inf_{k=1,\dots,4, \forall l, \forall t} \Phi^k_{l,t} + \inf_{k=5,6, \forall i, \forall t} \Phi^k_{i,t} \label{obj:DSO_attack} \\
        & \text{s.t.} 
        && \sum_{l \in \mathcal{L}} pf_{l,t} \cdot {\mathrm{LT}_{l,i}} - \sum_{l \in \mathcal{L}} pf_{l,t} \cdot {\mathrm{LF}_{l,i}} = \nonumber \\
&&& \begin{cases} 
P_{i,t}^d - p_{i,t}^{{g}*} (1 - y_{i,t}),  \text{if } i \in \mathcal{I}^{g}, \\
P_{i,t}^d - p_{sub,t}^{{g}}, \text{if } i \text{ is the substation}\\
P_{i,t}^d, ~ \text{if } i \notin \mathcal{I}^g \cup \{ \text{substation} \}
\end{cases}
        \label{cons:DSO_attack_P_flow} \\
&&& \sum_{l \in \mathcal{L}} qf_{l,t} \cdot {\mathrm{LT}_{l,i}} - \sum_{l \in \mathcal{L}} qf_{l,t} \cdot {\mathrm{LF}_{l,i}} = \nonumber \\
&&& \begin{cases} 
Q_{i,t}^d - q_{i,t}^{{g}*} (1 - y_{i,t}),  ~\text{if } i \in \mathcal{I}^{g}, \\
Q_{i,t}^d - q_{sub,t}^{{g}},  ~ \text{if } i \text{ is the substation},\\
Q_{i,t}^d, ~ \text{if } i \notin \mathcal{I}^g \cup \{ \text{substation} \}
\end{cases}
\label{cons:DSO_attack_Q_flow} \\
        & 
        && v_{\mathrm{FN}_{l},t} - v_{\mathrm{TN}_l,t} = 2 (r_{l} \cdot pf_{l,t} + x_{l} \cdot qf_{l,t}), \label{cons:DSO_attack_voltage} \\
         &&&   p^{g, min}_i \leq p_{i,t}^{g} \leq  p^{g, max}_i, q^{g, min}_i \leq q_{i,t}^{g} \leq  q^{g, max}_i,  \\
        & 
        && \Phi^1_{l,t} = pf_{l,t} - {PF}^{\mathrm{max}},~ \Phi^2_{l,t} = {PF}^{\mathrm{min}} - pf_{l,t},  \label{cons:sup_P_line_min_bound} \\
        & 
        && \Phi^3_{l,t} = qf_{l,t} - {QF}^{\mathrm{max}},~  \Phi^4_{l,t} = {QF}^{\mathrm{min}} - qf_{l,t},  \label{cons:sup_Q_line_min_bound} \\ 
        & 
        && \Phi^5_{i,t} =  v_{i,t}  - V^{\mathrm{max}}_i, ~ \Phi^6_{i,t} = V^{\mathrm{min}}_i -  v_{i,t} ,   \label{cons:sup_V_min_bound} \\
        & 
        && \| {y}_{t}\|_1   \leq K, \quad \forall i \in \mathcal{I},  \forall l \in \mathcal{L}, \forall t \in \mathcal{T}. \label{con:N-K_constraint}
    \end{align}
\end{subequations}
Here \( p^g_{\mathrm{sub},t} \) represents the power purchased from the substation at node \( i \), with an associated cost \( C^g_{\mathrm{sub},t} \). We define \( \mathcal{I}^{g}_a \) as the set of attacked generators. Constraints \eqref{cons:DSO_attack_P_flow}-\eqref{cons:DSO_attack_Q_flow} enforce active and reactive power flow balances at each node, considering active power generation from attackable DGs ($\mathcal{I}^{g}_a$), non-attackable DGs ($\mathcal{I}^{g} \setminus (\mathcal{I}^{g}_a \cup \text{sub})$), and the substation. DG values are set in Stage 1 and fixed in Stage 2, while substation power purchases remain adjustable in Stage 2 to satisfy the power balance.
An adversarial attack vector at each time \( t \), represented by \( y_t \), has a sum that must not exceed \( K \). This vector can attempt to push inequality constraints \eqref{cons:DSO_line_stat} and  \eqref{cons:DSO_voltage_bounds} beyond safe limits. Specifically, we relax constraints on active power flow (\( pf \)), reactive power flow (\( qf \)), and squared node voltage magnitude (\( v \)) in \eqref{cons:DSO_line_stat} and \eqref{cons:DSO_voltage_bounds} by terms $\Phi^k_{l,t}$ for \( k=1, \dots, 4 \) and $\Phi^k_{i,t}$ for \( k=5, \dots, 6 \) in \eqref{cons:sup_P_line_min_bound} - \eqref{cons:sup_V_min_bound}.

\subsection{Stage 3: Attack Mitigation} \label{sec: Attack_min}

Stage 3 focuses on mitigating the impacts of worst-case attacks identified in Stage 2 by deploying fast-response ESSs. The objective in this stage is to minimize the impact of constraint violations caused by these attacks, leveraging ESS to promptly maintain system safety.

Model \eqref{mod:Min_attack_storage} optimizes ESS dispatch to restore safety constraints well within safe bounds, while objective \eqref{obj:min_attack_storage} minimizes total system costs and reduces constraint violations (mitigation stage) caused by the attacks through strategic adjustments of the storage units. The first term represents the cost of energy supplied by storage units, while the second term represents the cost of power supplied by the substation, which has a higher cost parameter \( C_{i,t}^{\mathrm{ess}} \) than that of the DGs. Although unaffected by attacks, the substation power is used as part of the overall response to meet system demand. The reminding term, \( \sup[g_{\mathrm{cont}}(x^*, y^*, z)] \), corresponds to \( \sup_{k=1,\dots,4, \forall l, \forall t} \Phi^k_{l,t} + \sup_{k=5,6, \forall i, \forall t} \Phi^k_{i,t} \)  . Here, the supremum is used to enforce even the most critical constraints into the safe region, minimizing the number of constraints at risk of infeasibility and guiding the system toward stability. 
\begin{subequations}
    \label{mod:Min_attack_storage}
        \begin{align}
            &\min_{ {z}, {p}^g_{{sub}}}  
            && \sum_{i \in \mathcal{I}^{\mathrm{ess}}} \sum_{t \in \mathcal{T}} C_{i,t}^{\mathrm{ess}} p_{i,t}^{\mathrm{ess}} + \sum_{t \in \mathcal{T}} C^g_{\mathrm{sub},t}  p^g_{\mathrm{sub},t}  \notag\\
            & && + \sup_{k=1,\dots,4, \forall l, \forall t} \Phi^k_{l,t} + \sup_{k=5,6, \forall i, \forall t} \Phi^k_{i,t} 
            \label{obj:min_attack_storage} \\
            &\text{s.t.} 
            && \eqref{cons:DSO_attack_voltage} \sim \eqref{cons:sup_V_min_bound}, \\
&&& \sum_{l \in \mathcal{L}} pf_{l,t} \cdot {\mathrm{LT}_{l,i}} - \sum_{l \in \mathcal{L}} pf_{l,t} \cdot {\mathrm{LF}_{l,i}} = \nonumber \\
&&& \begin{cases} 
P_{i,t}^d - p_{i,t}^{{g}*} (1 - y_{i,t}^*) - \delta_i p^{\mathrm{ess}}_{i,t}, ~ \text{if } i \in \mathcal{I}^g, \\
P_{i,t}^d - p_{sub,t}^{{g}}, ~ \text{if } i \text{ is the substation}, \\
P_{i,t}^d, ~ \text{if } i \notin \mathcal{I}^g \cup \{ \text{substation} \},
\end{cases} \label{cons:storage_P_flow}\\
      & && \sum_{l \in \mathcal{L}} qf_{l,t} \cdot {\mathrm{LT}_{l,i}} - \sum_{l \in \mathcal{L}} qf_{l,t} \cdot {\mathrm{LF}_{l,i}} = \nonumber \\
&&& \begin{cases} 
Q_{i,t}^d - q_{i,t}^{{g}*} (1 - y_{i,t}^*) - \delta_i p^{\mathrm{ess}}_{i,t}, ~ \text{if } i \in \mathcal{I}^g, \\
Q_{i,t}^d - q_{sub,t}^{{g}}, ~ \text{if } i \text{ is the substation}, \\
Q_{i,t}^d, ~ \text{if } i \notin \mathcal{I}^g \cup \{ \text{substation} \},
\end{cases}\label{cons:storage_Q_flow} \\
            & &&  p^{\mathrm{ess}}_{i,t} =  p^ {\mathrm{dis}}_{i,t} - p^{\mathrm{ch}}_{i,t},  ~ \forall i \in \mathcal{I}^{\mathrm{ess}},   \label{cons:P_storage} \\
            & && \mathrm{soc}_{i,t} = \mathrm{soc}_{i, t-1} + \eta^{\mathrm{ch}} \frac{p^{\mathrm{ch}}_{i,t}}{E^{\mathrm{max}}}  
            - \frac{p^{\mathrm{dis}}_{i,t}}{\eta^{\mathrm{dis}} E_{\mathrm{max}}},   \forall i \in \mathcal{I} ^ {\mathrm{ess}},  \label{con:SOC_equality_eq} \\ 
            & && \beta_{i,t} ^ {\mathrm{ch}}\cdot p^{\mathrm{ch}}_{\mathrm{min}} \leq p^{\mathrm{ch}}_{i,t} \leq \beta_{i,t} ^ {\mathrm{ch}} \cdot p^{\mathrm{ch}}_{\mathrm{max}},  \forall i \in \mathcal{I}^{\mathrm{ess}},  \label{cons:P_Ch_bounds} \\
            & &&\beta_{i,t} ^ {\mathrm{dis}} \cdot p^{\mathrm{dis}}_{\mathrm{min}} \leq p^{\mathrm{Disch}}_{i,t} \leq \beta_{i,t} ^ {\mathrm{dis}} \cdot p^{\mathrm{dis}}_{\mathrm{max}},  \forall i \in \mathcal{I}^{\mathrm{ess}} \label{cons:P_Disch_bounds} \\
            & && \mathrm{soc}^{\mathrm{min}} \leq \mathrm{soc}_{i,t} \leq \mathrm{soc}^{\mathrm{max}},   \forall i \in \mathcal{I}^{\mathrm{ess}} . \label{cons:SOC_limits}\\
            & &&\beta_{i,t} ^ {\mathrm{ch}} + \beta_{i,t} ^ {\mathrm{dis}} \leq 1 ,~  \beta_{i,t} ^ {\mathrm{ch}}, \beta_{i,t} ^ {\mathrm{dis}} \in \mathrm{binary}~ \{0, 1\}  \label{cons:batterystatue}
        \end{align}
\end{subequations}
The notation in   \eqref{cons:storage_P_flow} - \eqref{cons:storage_Q_flow} can be interpreted as follows:
\begin{enumerate}
    \item \textbf{Case 1:} For \( i \in \mathcal{I}^g \), \( y_{i,t}^* \) is the fixed attack variable from Stage 2, adjusting generation through \( p_{i,t}^{g*} (1 - y_{i,t}^*) \). For unattacked generators \( i \in \mathcal{I}_a \setminus \mathcal{I}_a^g \) (i.e., \( y_{i,t}^* = 0 \)), generation remains unaltered. \( \delta_i \) is the ESS indicator, set to \( 1 \) if \( i \in \mathcal{I}^{\text{ess}} \) and \( 0 \) otherwise.

    \item \textbf{Case 2:} For the substation, \( P_{i,t}^d - p_{sub,t}^{g} \) applies, without ESS or attack variables.

    \item \textbf{Case 3:} For nodes not in \(\mathcal{I}^g\) and not the substation, \( P_{i,t}^d \) is used, as there is no generation, ESS, or attack associated with these nodes.
\end{enumerate}
The battery status constraint \eqref{cons:batterystatue} with binary variables \( \beta_{i,t}^{\mathrm{ch}}, \beta_{i,t}^{\mathrm{dis}} \) prevents simultaneous charging and discharging. The SOC is updated in \eqref{con:SOC_equality_eq}, accounting for charging/discharging efficiency. Power limits for charging and discharging are set by \eqref{cons:P_Ch_bounds} and \eqref{cons:P_Disch_bounds}. The SOC limits are enforced in \eqref{cons:SOC_limits} to ensure ESSs maintain adequate reserve capacity, enabling effective responses to cyberattacks and mitigating system disruptions.

\section{Experimental Results}

Our model was evaluated on the IEEE 33-node test system, as shown in Fig. \ref{fig:test_syst}. In this network, nodes 4, 10, 13, 18, 25, 27, and 33 function as DG nodes, with node 13 specifically hosting a PV-based renewable source. The system connects to the main grid at node 1. ESS units are deployed at nodes 4, 10, 18, 25, 27, and 33 to support the system under attack scenarios if the DGs are impacted. 
The optimization problem was solved using GUROBI 11.0.3 and IPOPT 3.14.11 on the Pyomo 6.5.0 platform.
\begin{figure}[htpb!]
    \centering
    \includegraphics[width=1\linewidth]{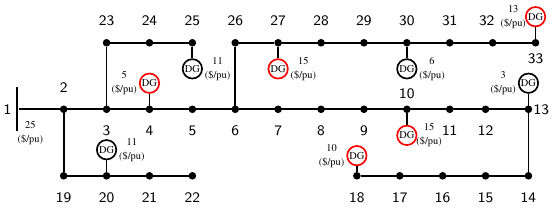}
    \caption{The 33-node DS test system; $\bullet$: nodes; \tikzmarknode[mycircled,draw=black,fill=white, font=\tiny]{t1}{\textcolor{white}{1}}: non-attackable DG nodes; \tikzmarknode[mycircled,draw=red,fill=white, font=\tiny]{t1}{\textcolor{white}{1}}: attackable DG nodes. }
    \label{fig:test_syst}
    \vspace{-0.6cm}
\end{figure}
\begin{figure}[htp!]
    \centering
    \includegraphics[width=0.9\linewidth]{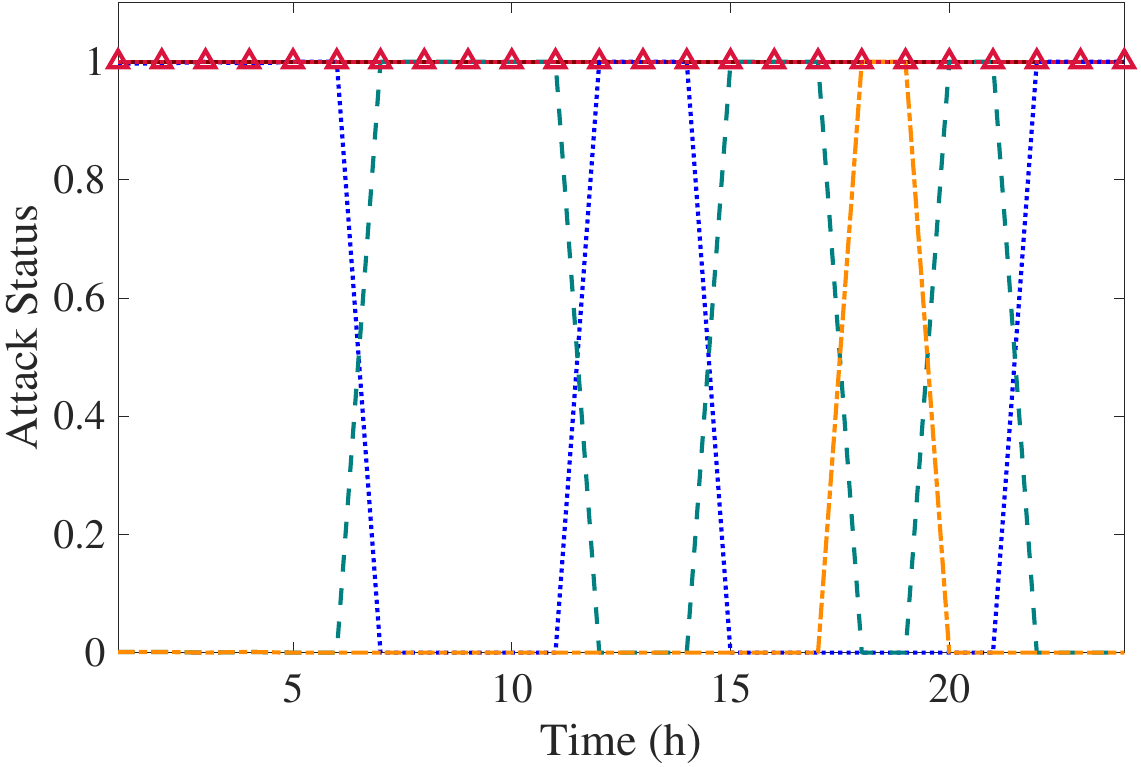}
    \caption{
        Attack status over different nodes; Nodes:
        \textcolor[rgb]{0.5, 0.0, 0.0}{\textbf{---}} Node 4, 
        \textcolor[rgb]{0.0, 0.0, 1.0}{\textbf{$\boldsymbol{\cdots}$}} Node 10, 
        \textcolor[rgb]{0.0, 0.5, 0.5}{\textbf{-- --}} Node 27, 
        \textcolor[rgb]{1.0, 0.55, 0.0}{\textbf{-- - --}} Node 33,  
        \textcolor[rgb]{0.86, 0.08, 0.24}{\textbf{$\boldsymbol{\cdots  }$}\scriptsize{\textbf{$\boldsymbol{\bigtriangleup}$}}} Node 18.
    }
    \label{fig:Attack_status}
    \vspace{-0.5cm}
\end{figure}
\subsection{Adversarial Attack Analysis}
This subsection examines the vulnerability and attack status of DG units under adversarial attacks in the 33-node test system. Attack statuses range from 0 (no attack) to 1 (fully active attack), with intermediate values indicating partial attacks.
In Fig. \ref{fig:Attack_status}, Nodes 4 and 18 experienced sustained attacks, indicating high vulnerability. Node 10 faced intermittent attacks, with full attacks interspersed with no attacks, especially at midday and late. Node 27 had brief early attacks, followed by full attacks, then no attacks. Node 33 had minimal early partial attacks, followed by a full attack later, and then no attack. These patterns indicate the need for tailored defense strategies to address varying attack levels, underscoring the importance of timing in effective mitigation efforts.

\begin{figure}[htb!]
\vspace{-0.4cm}
    \centering
    \includegraphics[width=0.9\linewidth]{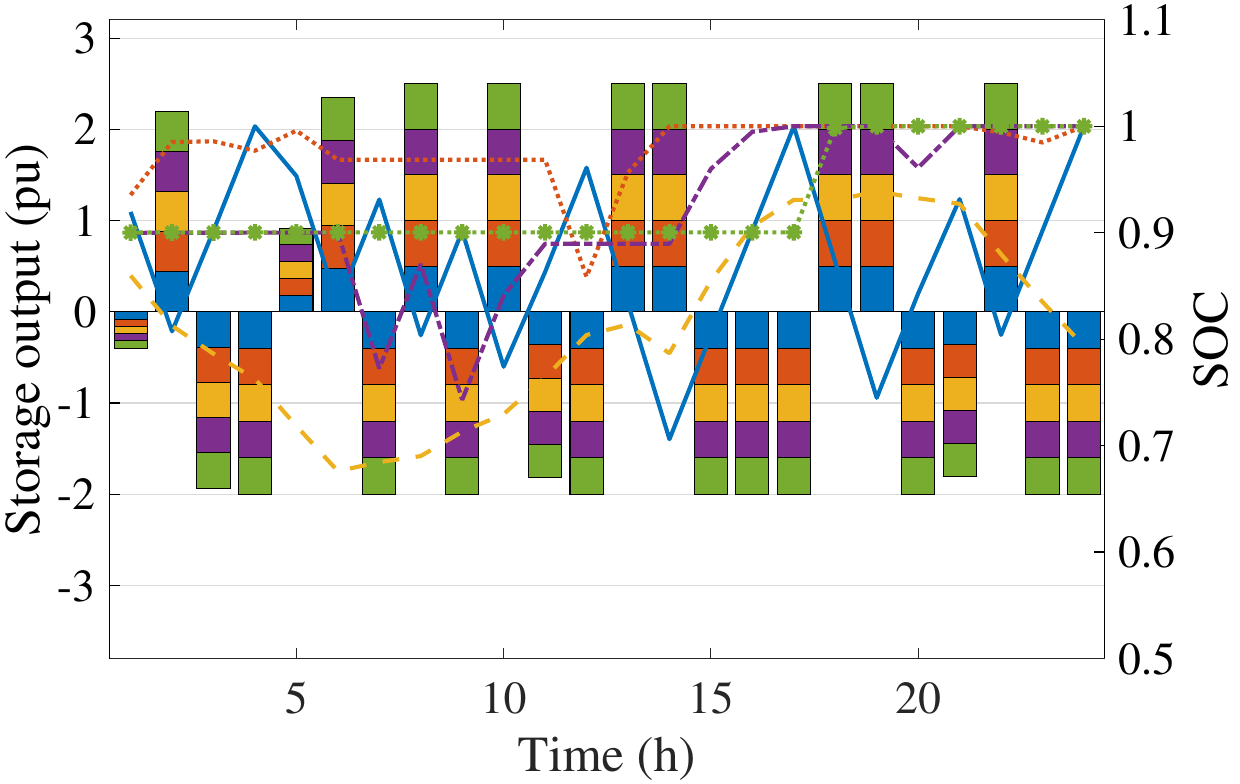}
    \caption{Storage generation output and their SOC over 24 hours; Storage output nodes: \Rectangle[storageBlue] 4, \Rectangle[storageOrange] 10, \Rectangle[storageYellow] 18, \Rectangle[storagePurple] 27, \Rectangle[storageGreen] 33. Storage SOC: nodes:  \textcolor{storageBlue}{\textbf{---}} 4, $\textcolor{storageOrange}{\boldsymbol{\cdot \cdot\cdot} }$ 10, \textcolor{storageYellow}{\textbf{-- --}} 18, \textcolor{storagePurple}{\textbf{-- - --}} 27, $\textcolor{storageGreen}{\boldsymbol{\cdot \cdot *} }$ 33 }
    \label{fig:Storage_out}
    \vspace{-0.5cm}
\end{figure}

\subsection{ESS Deployment for Attack Mitigation}
Following worst-case attack identification, energy storage was deployed at vulnerable nodes for impact mitigation. Fig. \ref{fig:Storage_out} displays charge, discharge, and SOC levels for each storage unit over 24 hours, with positive values for discharges, negative for charges, and SOC shown as lines.
A comparison of Figs. \ref{fig:Attack_status} and \ref{fig:Storage_out} show that energy storage deployment during attacks significantly supported the network. Sustained attacks on nodes 4, 18, and 33 led to active use of storage units, with Node 4 showing alternating charge-discharge cycles and Node 18's SOC fluctuating to counteract attacks.
Nodes 10 and 27 maintained high SOC with minimal discharge, serving as backup capacity, while Node 33 acted as a reserve with limited SOC variation. These findings underscore the importance of coordinated energy management for ESSs to maintain sufficient capacity for future network resilience.

\begin{figure*}[htpb!]
    \centering
    \begin{subfigure}{0.32\textwidth}
        \centering
        \includegraphics[width=\textwidth]{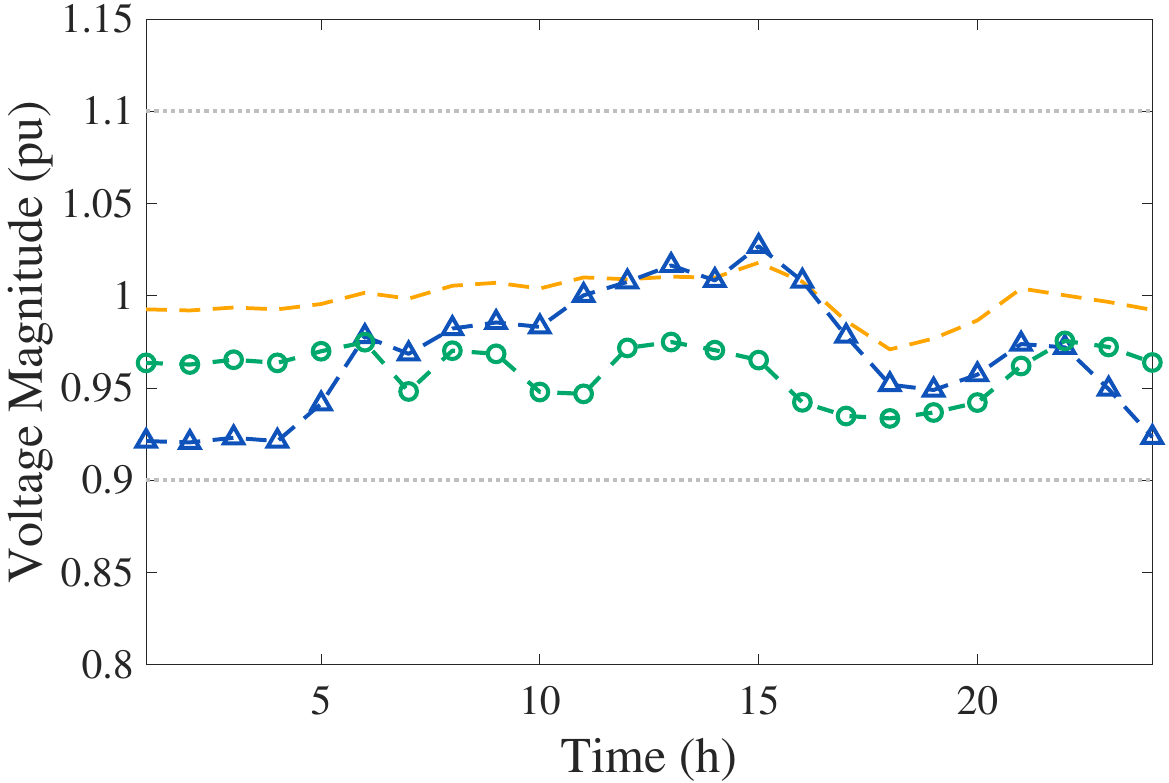}
        \caption{Stage 1}
        \label{fig:voltage_stage1}
    \end{subfigure}
    \hfill
    \begin{subfigure}{0.32\textwidth}
        \centering
        \includegraphics[width=\textwidth]{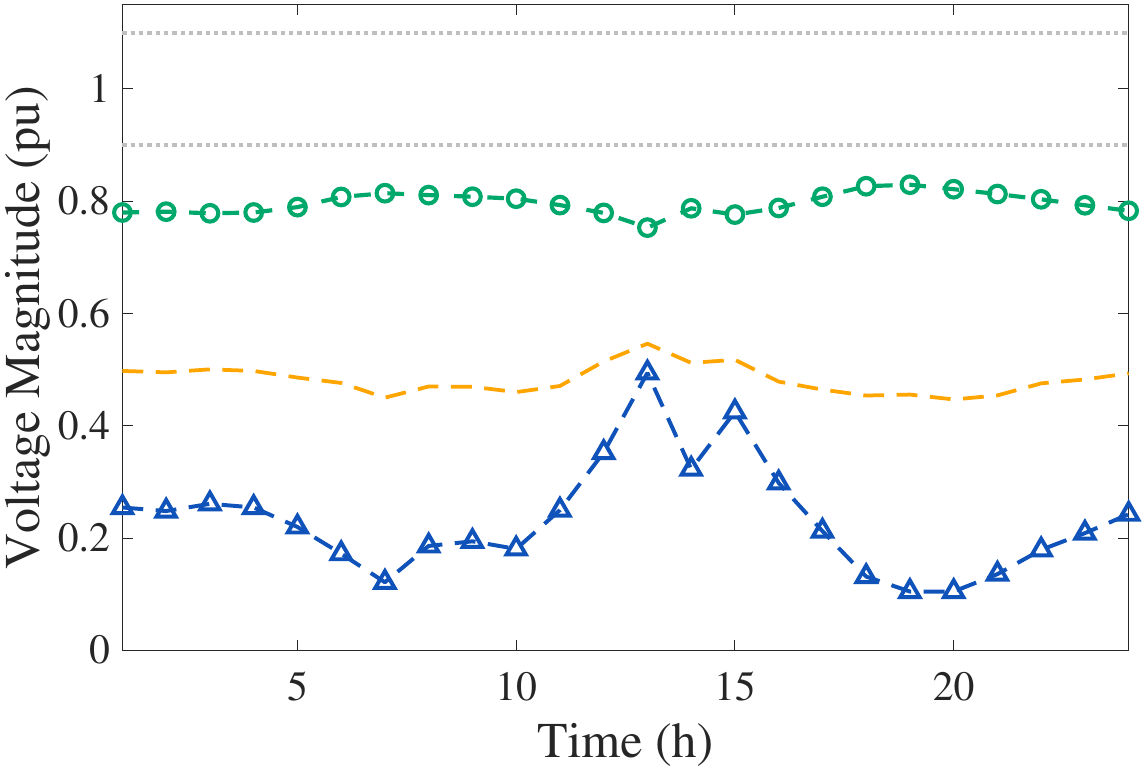}
        \caption{Stage 2}
        \label{fig:voltage_stage2}
    \end{subfigure}
    \hfill
    \begin{subfigure}{0.32\textwidth}
        \centering
        \includegraphics[width=\textwidth]{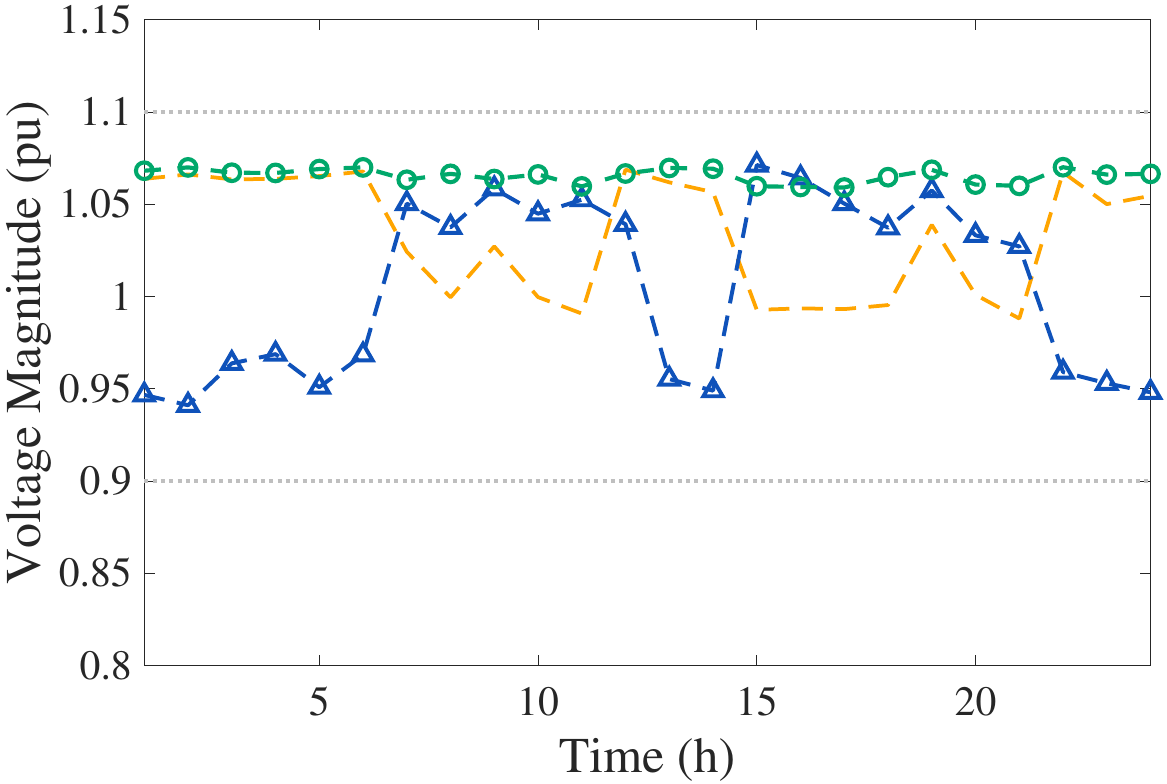}
        \caption{Stage 3}
        \label{fig:voltage_stage3}
    \end{subfigure}
    \captionsetup{font=small}
    \vspace{-0.1cm}
    \caption{
        Voltage magnitudes for nodes across different stages over 24 hours; Nodes:
        \textcolor{orange}{\textbf{-- - --}} Node 6,
        \textcolor{blue}{\textbf{-- --}{\scriptsize{$\boldsymbol{\bigtriangleup}$}}} Node 10,  
        \textcolor{greenJ}{\textbf{-- --$\boldsymbol{\circ}$}} Node 24. 
        Boundary limits: \textcolor{lightgray}{\textbf{$\boldsymbol{\cdots}$}} minimum (0.9 pu), \textcolor{lightgray}{\textbf{$\boldsymbol{\cdots}$}} maximum (1.1 pu).
}

    \label{fig:voltage_comparison}
    \vspace{-0.5cm}
\end{figure*}

\begin{figure*}[htb!]
    \centering
    \begin{subfigure}{0.32\textwidth}
        \centering
        \includegraphics[width=\textwidth]{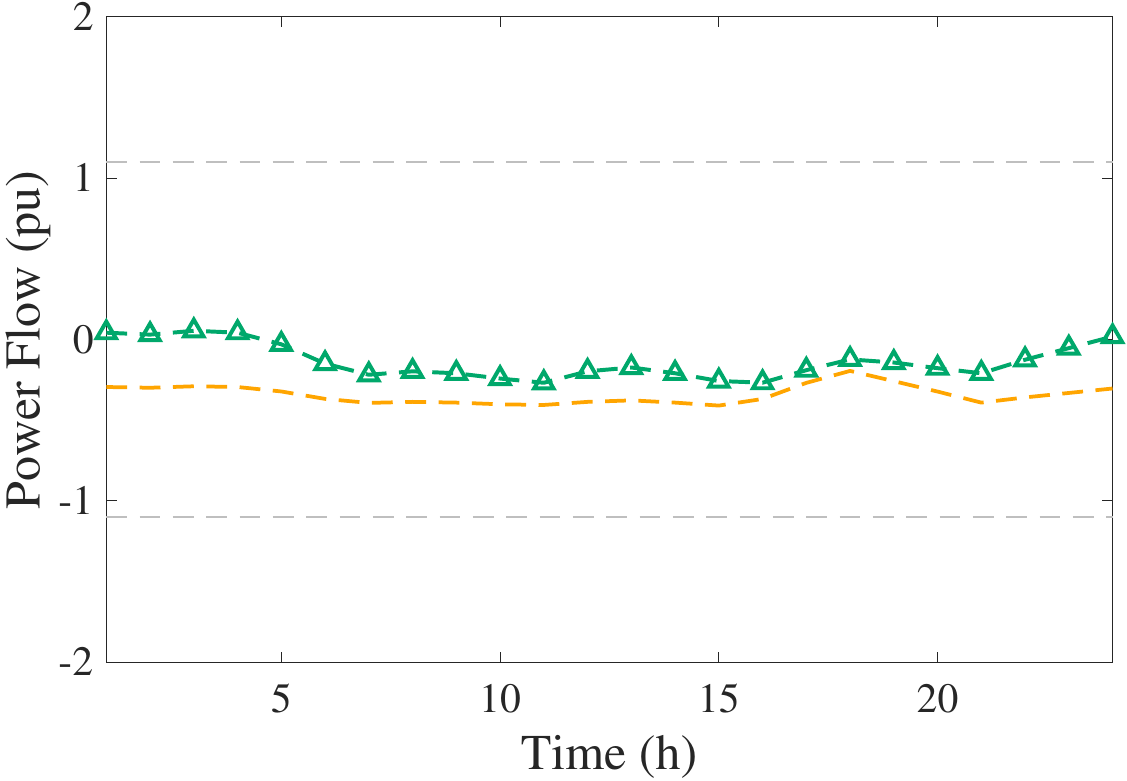}
        \caption{Stage 1}
        \label{fig:powerflow_stage1}
    \end{subfigure}
    \hfill
    \begin{subfigure}{0.32\textwidth}
        \centering
        \includegraphics[width=\textwidth]{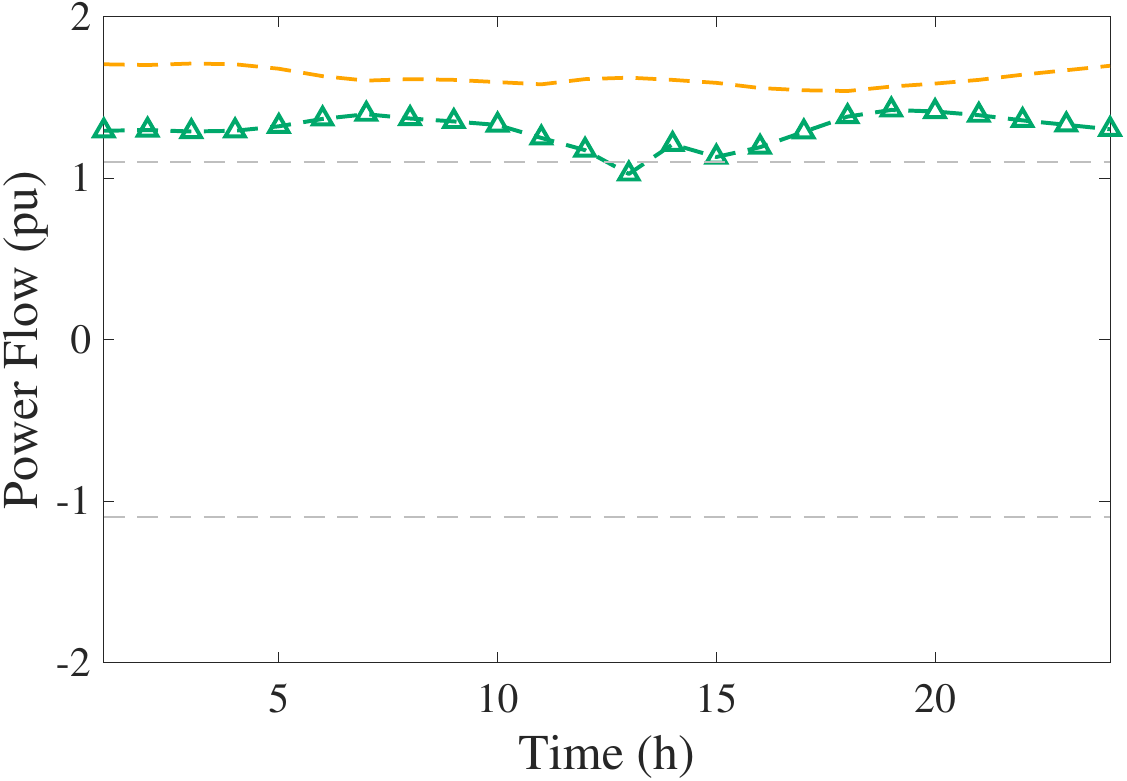}
        \caption{Stage 2}
        \label{fig:powerflow_stage2}
    \end{subfigure}
    \hfill
    \begin{subfigure}{0.32\textwidth}
        \centering
        \includegraphics[width=\textwidth]{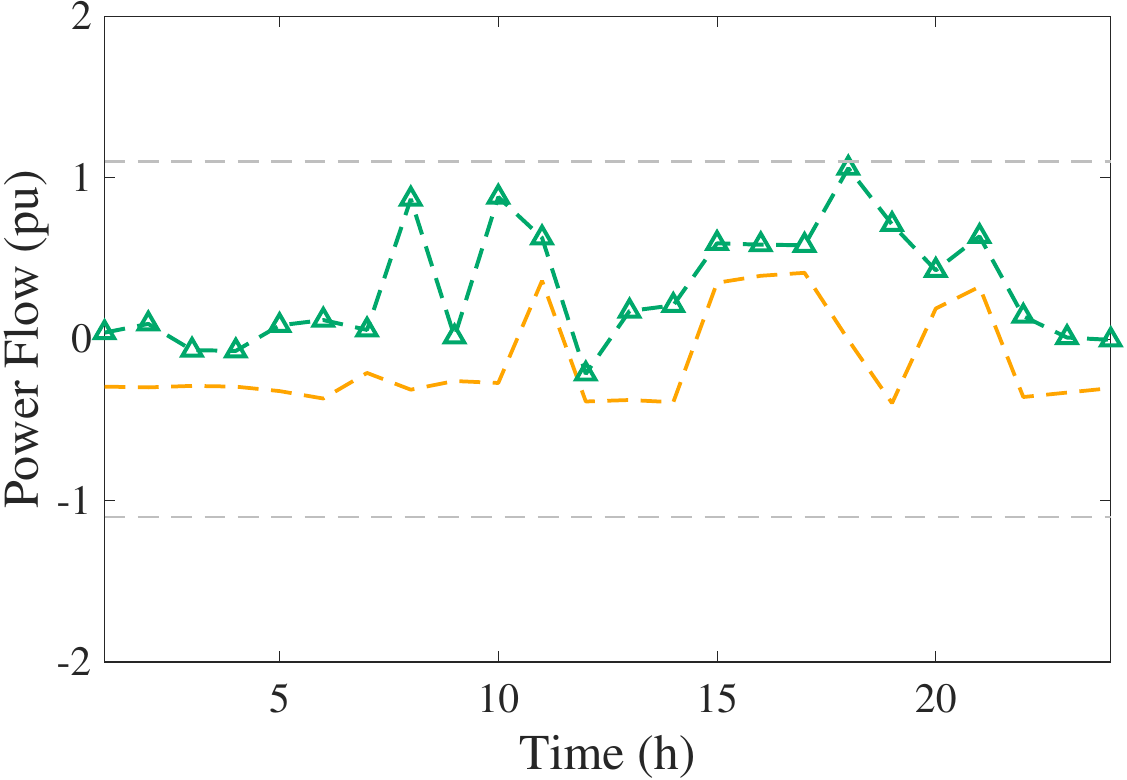}
        \caption{Stage 3}
        \label{fig:powerflow_stage3}
    \end{subfigure}
    \captionsetup{font=small}
        \vspace{-0.1cm}
    \caption{
        Power flow for Lines 2 and 4 across different stages over 24 hours; Lines:
        \textcolor{orange}{\textbf{-- - --}} Line 2,
        \textcolor{greenJ}{\textbf{-- --}{\scriptsize{$\boldsymbol{\bigtriangleup}$}}} Line 4   
        Boundary limits: \textcolor{lightgray}{\textbf{$\boldsymbol{\cdots}$}} minimum (-1.5 pu), \textcolor{lightgray}{\textbf{$\boldsymbol{\cdots}$}} maximum (1.5 pu).
    }
    \label{fig:power_flow}
    \vspace{-0.7cm}
\end{figure*}

\subsection{System Parameter Analysis}

This section evaluates the impact of Stage 2 attacks in severely disrupting power systems and causing potential outages, as well as the effectiveness of Stage 3 mitigation strategies in successfully restoring the system to normal conditions.

\subsubsection{Voltage Magnitude Analysis}
The performance of Volt-Var control, essential for maintaining power system stability, was analyzed at nodes 6, 10, and 24 (Fig. \ref{fig:voltage_comparison}), demonstrating the impact of worst-case adversarial attacks and the effectiveness of Stage 3 mitigation strategies in restoring normal conditions. Under normal conditions (Fig. \ref{fig:voltage_stage1}), voltage remained stable within the 0.9 to 1.1 pu range. During attacks (Fig. \ref{fig:voltage_stage2}), voltage dropped sharply, with Node 6 reaching as low as 0.45 pu, illustrating the severe stress imposed by the attackers. Stage 3 mitigation with ESS (Fig. \ref{fig:voltage_stage3}) effectively restored voltage magnitudes, maintaining nodes 6, 10, and 24 within the desired 0.9 to 1.1 pu range and countering attack effects, particularly at Node 10, thereby ensuring power system safety.

\subsubsection{Power Flow Analysis}
Power flow analysis was conducted to evaluate the potential for outages due to cyberattacks and the system’s ability to recover. Lines 2 and 4 were selected as examples for this assessment (Fig. \ref{fig:power_flow}). Under normal conditions (Fig. \ref{fig:powerflow_stage1}), power flows on both lines remained within safe operational limits. However, during the attack phase (Fig. \ref{fig:powerflow_stage2}), power flows on both lines exceeded the 1.1 pu threshold, with Line 2 approaching critical levels and risking system instability. In Stage 3 (Fig. \ref{fig:powerflow_stage3}), ESS deployment effectively mitigated these adverse effects, restoring Line 2 to within safe operational limits and stabilizing Line 4. These findings highlight the crucial role of ESS in maintaining system safety and resilience in the face of adverse conditions.

\section{Conclusion}
This paper introduces a tri-level optimization framework to enhance power system resilience against adversarial attacks by integrating economic dispatch, vulnerability assessment, and mitigation through ESS. The first stage optimizes DG operation and power supply costs from substations under normal conditions, the second stage identifies vulnerabilities in worst-case attack scenarios, and the third stage leverages ESS to restore system safety and enhance resilience. Evaluation of the IEEE 33-node system demonstrates effective maintenance of safe voltage levels and power flow during attacks, highlighting the role of ESS in supporting resilience and real-time energy management. Future work may focus on scaling the framework to larger networks and exploring advanced mitigation strategies through safe reinforcement learning.

\begin{footnotesize}

\end{footnotesize}
\end{document}